\documentclass[runningheads]{llncs}
\usepackage{booktabs}
\usepackage{multirow}
\usepackage{amsfonts}
\usepackage{mathrsfs}
\usepackage{amssymb}
\usepackage{dsfont}
\usepackage{scrextend}
\usepackage{mathtools}
\usepackage{graphicx}
\usepackage{xparse} 
\usepackage{tikz}
\usetikzlibrary{calc,positioning,fit,shapes,shapes.geometric,arrows.meta,calc,trees,patterns}

\tikzset{
  atom/.style={draw, thick, circle, minimum size=2mm,inner sep=0pt, semitransparent},
  compart/.style={draw, very thick, nearly transparent, rectangle,
    rounded corners},
  source/.style={draw,thick,rounded corners,fill=yellow!20,inner sep=.3cm, minimum width=3cm},
  process/.style={source,fill=green!20, minimum width = 3cm},
  arrow/.style={-{Stealth[]}, thick,},
  lab/.style={draw=none, font=\bfseries},
  every node/.style={align=center},
}

\DeclareDocumentCommand{\ClusterBox}{O{yellow} m m m O{0} m O{0}}{
\begin{scope}[shift={(#5,#7)}]
  \node[compart,fill=yellow,minimum width=#4cm, minimum
  height=#6cm,inner sep=1mm,anchor=west, path picture={
  \foreach \i in {1,2,...,#3}
  {
    \path let \p1=(path picture bounding box.south west),
    \p2=(path picture bounding box.north east),
    \n1={rnd}, \n2={rnd} in
    ({\n1*\x1+(1-\n1)*\x2},{\n2*\y1+(1-\n2)*\y2}) node[atom, fill=#1](#2-atom\i){};
  }
}] at (0,0) (#2-comp) {};
\node[atom, fill=#1, above left=3mm and 2mm of #2-comp.south east] (#2-rightatom) {};
    
\end{scope}
}

\DeclarePairedDelimiter{\ceil}{\lceil}{\rceil}
\newcommand{\happyend}{\ensuremath{\mathcal{S}_3}}
\newcommand{\extpred}{\ensuremath{\mathcal{S}_2}}
\newcommand{\extall}{\ensuremath{\mathcal{S}_1}}

\begin{document}
\title{SPSC: a new execution policy for exploring discrete-time stochastic simulations
}
\titlerunning{SPSC: A new execution policy for stochastic simulations}
%
\author{
Yu-Lin Huang \and Gildas Morvan \and Fr\'ed\'eric Pichon \and David Mercier
}
\authorrunning{
Y-L. Huang et al.
}
%
\institute{
Univ. Artois, EA 3926, Laboratoire de G\'enie Informatique et d'Automatique de l'Artois (LGI2A), B\'ethune, France.\\ \email{\{ylin.huang, firstname.lastname\}@univ-artois.fr}
}
\maketitle              
\begin{abstract}
In this paper, we introduce a new method called SPSC (Simulation, Partitioning, Selection, Cloning) to 
estimate efficiently the probability of possible solutions in stochastic simulations. This method can be applied to any type of simulation, however it is particularly suitable for multi-agent-based simulations (MABS). Therefore, its performance is evaluated on a well-known MABS and compared to the classical approach, \emph{i.e.}, Monte Carlo.   

\end{abstract}
\section{Introduction}
\label{sec:intro}

\begin{figure}[htbp]
\centering \scalebox{.55}{ \begin{tikzpicture}
  \node[draw, thick, circle] at (0,0) (model) {Stochastic\\Model}; 
  \node[above left= .1em and 6em of model] (IC) {Initial conditions};
  \node[below left= .1em and 6em of model] (PS) {Parameter set};
  \node[right= 6em of model] (OB) {Observables};

  \draw[arrow] (IC.east) -- (model.west);  
  \draw[arrow] (PS.east) -- (model.west);
  \draw[arrow] (model.east) -- (OB.west);
  \draw[arrow] (model) to [out=255,in=285,looseness=5] node[below] {$\Delta t$} (model);

\end{tikzpicture} }
\caption{Illustration of a discrete-time stochastic simulation}
\label{fig:stochastic}
\end{figure}
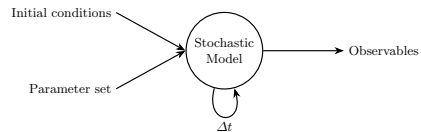

Multi-agent-based simulations (MABS) are widely used in various fields to study complex systems~\cite{Railsback:2011}. Most of them are combined with stochasticity to represent non fully controllable phenomena and use a discrete-time approach to facilitate model construction.  Such model can generally be described as taking some initial conditions and some parameter set as inputs, in order to return outputs at each time step (\emph{c.f.} Figure \ref{fig:stochastic}). 

Before 
running into exploration of the parameter set or the initial condition  space, we  
must 
first analyze  outcomes from a fixed parameter set and initial conditions.
Let us denote a stochastic simulation outputs (called observables in the following) at a final time step $T$ as a random vector $\boldsymbol{X}_{T}$. Then a key question to address is: 
what is the probability $\mathbb{P}(\boldsymbol{X}_{T} \in \mathcal{S}) = \theta_{\mathcal{S}}$ of a specific solution $\mathcal{S}$?

The classical method to handle this question is Monte Carlo simulation (MC)~\cite{Rubinstein:2016}. It consists in simulating a number $n$ of replications and building an estimator $\hat{\theta}_\mathcal{S}$ of $\theta_{\mathcal{S}}$ defined as: 
\begin{align}
\label{eq:rel_freq}
\hat{\theta}_{\mathcal{S}} = \frac{1}{n} \sum_{i=1}^n  \mathds{1}_{\mathcal{S}}(\boldsymbol{X}_{T}^i)
\end{align}
where $\mathds{1}_{\mathcal{S}}$ is the indicator function of the set $\mathcal{S}$ and $\boldsymbol{X}_{T}^i$ the value of observables in the $i^{\text{th}}$ replication. The issue with this approach is that for the estimator to be good, the number $n$ has generally to be large, as illustrated in Section \ref{sec:MC}.  

Some methods have been developed to speed-up the computation of such simulations, such as splitting~\cite{LEcuyer:2009} or polyagent~\cite{Parunak:2011}. However, they look for specific solutions (rare or mean), assume a particular modeling approach (Markov chains or agent-based) and require some low-level manipulations of the model. 

In this paper, we propose a policy that  simulates an authorized number $N$ of replications and is as generic as the MC approach yet provides a better estimator when computational resources are  limited (\emph{i.e.} small $N$).

The paper is organized as follows: Section \ref{sec:MC} recalls and illustrates a standard approach to determine the required number of replications in Monte Carlo simulation for a single observable. The design principles and the approach proposed to answer the above-mentioned issues are presented in Section~\ref{sec:proposition} and then applied to a classical MABS in Section~\ref{sec:experiences}. Section~\ref{sec:conclusion} concludes the paper.

\section{Monte Carlo simulation, how many replications?}
\label{sec:MC}
 We recall in this section a standard approach to determine the number $n(X_{T,i})$ of replications to obtain a good estimator $\hat{\theta}_{S_i}$ of the probability $\mathbb{P}(X_{T,i} \in S_i) = \theta_{S_i}$ of some solution $S_i$ where $X_{T,i}$ is one observable of the vector $\boldsymbol{X}_{T}$. Suppose a desired relative error $\epsilon$ for the estimator $\hat{\theta}_{S_i}$ at confidence level $1-\alpha$~:
 \begin{align}
     \label{eq:minimal_n}
     \mathbb{P}(\frac{|\hat{\theta}_{S_i} - {\theta}_{S_i}|}{ {\theta}_{S_i}} \le \epsilon) \ge 1-\alpha.
 \end{align}
 The minimal value for $n(X_{T,i})$ to verify (\ref{eq:minimal_n}) can be determined by applying the following algorithm~\cite[p. 449]{Banks:2010}:
\begin{enumerate}
    \item Simulate $n_0$ replications. ($n_0$ observations $X_{T,i}^1, \dots, X_{T,i}^{n_0}$)
    \item Compute
    \begin{equation}
    	n(X_{T,i}) = \ceil{
    	(\frac{Z_{1-(\frac{\alpha}{2})} \cdot s_i}{\epsilon \cdot \overline{X}_{T,i}})^2}
    	\label{eq:n}
    \end{equation}
    where 
    $Z_{1-(\alpha / 2)}$ is the $100(1-\alpha/2)$ quantile of the normal distribution, $s_i$ stands for the sample standard deviation over the $n_0$ observations and $\overline{X}_{T,i}$ is the sample mean value over the $n_0$ observations. The conventional values for $n_0$, $\epsilon$ and $\alpha$ are respectively $150$, $0.05$ and $0.05$.
\end{enumerate}
Afterward, we can then deduce the necessary number $n$ satisfying every observable as:
\begin{align}
    n = \max_{X_{T,i} \in \boldsymbol{X_T} } n(X_{T,i})
\end{align}

To illustrate this algorithm, let us take an academic example. We consider an environment containing vegetation and 2 types of agents: preys consuming the vegetation and predators hunting preys for food. Both preys and predators can move without restriction in the environment. This model has been implemented on the \textsc{Similar} platform~\cite{Morvan:2017} and is based on the NetLogo wolf sheep predation model~\cite{Wilensky:1997}. The set of observables here consists of the populations of different species at each time step. The necessary number $n$ for some arbitrarily chosen parameter set and initial state of the simulation,  using the conventional values for $n_0$, $\epsilon$ and $\alpha$, is 3600 (\emph{c.f.} Table \ref{tab:nb_simus}). However, if we want a more precise estimation, the necessary number $n$ of replications increases drastically: for example, considering a relative error $\epsilon=0.005$ yields a necessary number of replications $n = 7249285$. 

\begin{table}[htbp]
    \caption{Determination of the necessary number of replications for the prey predator model implemented on the \textsc{Similar} platform. The parameters applied are $n_0 = 150$, $\epsilon = 0.05$ and $\alpha = 0.05$.}
    \label{tab:nb_simus}
    \begin{center}
        \begin{tabular}{c@{\hspace{.4cm}}c@{\hspace{.4cm}}c@{\hspace{.4cm}}c}
            \toprule
            $X_{T,i}$&$s_i$&$\overline{X}_{T,i}$&$n(X_{T,i})$\\
            \midrule
            number of preys&697.83&783.77&1219\\
            number of predators&196.95&128.67&\textbf{3600}\\
            \bottomrule
        \end{tabular}
    \end{center}
\end{table}%

\section{A new execution policy for stochastic simulations}
\label{sec:proposition}
    In this section, we introduce a new execution policy for stochastic simulations called SPSC (Simulation, Partitioning, Selection, Cloning). This approach  relies on 
    a decomposition of the probability of interest 
    that we explain first.

	\subsection{Decomposition of the probability of interest}
	The probability $\mathbb{P}(\boldsymbol{X}_{T} \in \mathcal{S})$ concerns the observables with respect to a specific solution $\mathcal{S}$ at some final time step $T$. Thanks to the law of total probability, considering some intermediate time step $j$ before $T$, we can write 
	\begin{align}
	    \mathbb{P}(\boldsymbol{X}_{T} \in \mathcal{S}) = \sum_{\mathcal{S}_j \in \mathscr{P}_j}
	    \mathbb{P}(\boldsymbol{X}_{T} \in \mathcal{S} | \boldsymbol{X}_{j} \in \mathcal{S}_j)\mathbb{P}(\boldsymbol{X}_{j} \in \mathcal{S}_j)
	    \label{eq:decomp}
	\end{align}
	where $\mathscr{P}_j$ is a partition  of the state space of the random vector $\boldsymbol{X}_{j}$. 
	
	More generally, considering all time steps before $T$, 
	we can obtain the following decomposition by  assuming a discrete-time system where $\boldsymbol{X}_{i}$ depends only on $\boldsymbol{X}_{i-1}$:
	
		\begin{align}
	\label{eq:full_decomp}
	    \mathbb{P}(\boldsymbol{X}_{T} \in \mathcal{S}) = \sum_{
	        \substack{\mathcal{S}_{T-1} \in \mathscr{P}_{T-1}\\ 
	                  \vdots                                     \\
	                  \mathcal{S}_{1} \in \mathscr{P}_{1}
	        }} \prod_{i=0}^{T-1}
	    \mathbb{P}(\boldsymbol{X}_{i+1} \in \mathcal{S}_{i+1} | \boldsymbol{X}_{i} \in \mathcal{S}_{i})
	\end{align}
	where $\mathscr{P}_i$, $i=1,...,T-1$, is a partition of the state space of $\boldsymbol{X}_{i}$, 
	$\mathcal{S}_{T} =\mathcal{S}$ and $\mathcal{S}_{0}$ is the initial state of the simulation.
	\subsection{SPSC: Simulation, Partitioning, Selection, Cloning}
	
	
	
\begin{figure}[htbp]
    \centering \scalebox{.75}{ \begin{tikzpicture}
  \node[source] at (0,0) (start) {Initial states}; 
  \node[process, below = 1.5cm of start] (simu) {SIMULATION};
  \node[process, below = 1.5cm of simu ] (clust) {PARTITIONING};
  \node[process, left  = 1.5cm of clust] (select) {SELECTION};
  \node[process, above  = 1.5cm of select ] (clone) {CLONING};
  \node[source,  right = 2.5cm of simu ] (end) {Final states};

  \node[compart, fit=(simu) (clust) (select) (clone), inner sep = .7em, fill=blue!10] (loop) {};

  \node at (loop.center) (loop-text) {loop from $t_{(0)}$ to $t_{(m-1)} $};

  \draw[arrow] (start) -- node[midway, right] {enter\\at time $ t_{(0)}$} (simu);
  \draw[arrow] (simu) -- (clust);
  \draw[arrow] (clust) -- (select);
  \draw[arrow] (select) -- (clone);
  \draw[arrow] (clone) -- (simu);
  \draw[arrow] (simu) -- node[midway, below] {at time $t_{(m)}$} 
  	node [midway, above] {exit} (end);
\end{tikzpicture} }
    \caption{SPSC process diagram}
    \label{fig:diag_proc}
    \end{figure}
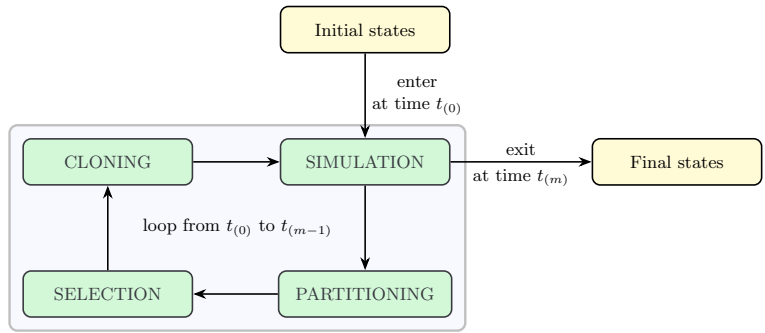

\begin{figure}[htbp]
    \centering \scalebox{.65}{ \begin{tikzpicture}
  \pgfmathsetseed{42}
  \node[atom] (start) at (-2,0) {};
  \ClusterBox[yellow!50]{clust1}{5}{1}[2]{1.5}[2]
  \ClusterBox[yellow!50]{clust2}{15}{1}[2]{2}[0]
  \ClusterBox[yellow!50]{clust3}{10}{1}[2]{1.5}[-2]



  \node[compart, fit = (clust1-comp) (clust2-comp) (clust3-comp), inner sep=2mm](T1) {};

  \node[lab, below = .3 of T1] (time-1) {$t_{(1)}$};
  \node[lab] at (start|-time-1) (time-0) {$t_{(0)}$};

  \foreach \angle in {105, 120, ..., 255}{
  \draw[arrow] (start.east) -- 
               (T1.\angle);
  }


  \node[atom, fill=blue] at (clust1-atom4) (delegate1) {};
  \node[atom, fill=blue] at (clust2-atom1) (delegate2) {};
  \node[atom, fill=blue] at (clust3-atom6) (delegate3) {};

  \node[atom, fill=blue, right = 3 of delegate1]  (selected1) {};
  \node[atom, fill=blue] at (selected1|-delegate2)  (selected2) {};
  \node[atom, fill=blue] at (selected1|-delegate3)  (selected3) {};


  \draw[arrow, dashed] (delegate1) -- node[pos=0.65, above, font=\bfseries]  {\scriptsize delegate} 
          node[pos=0.65, below, font=\bfseries]  {\scriptsize of $\mathcal{S}_{(1)}^1$} 
          (selected1);
  \draw[arrow, dashed] (delegate2) -- node[pos=0.65, above, font=\bfseries]  {\scriptsize delegate} 
          node[pos=0.65, below, font=\bfseries]  {\scriptsize of $\mathcal{S}_{(1)}^2$} 
          (selected2);
  \draw[arrow, dashed] (delegate3) -- node[pos=0.65, above, font=\bfseries]  {\scriptsize delegate} 
          node[pos=0.65, below, font=\bfseries]  {\scriptsize of $\mathcal{S}_{(1)}^3$} 
          (selected3);

  \ClusterBox[yellow!50]{clust21}{16}{1}[10]{1.5}[2]
  \ClusterBox[yellow!50]{clust22}{3}{1}[10]{1.5}[-2]
  \ClusterBox[yellow!50]{clust23}{12}{1}[10]{2}[0]

  \node[compart, fit = (clust21-comp) (clust22-comp) (clust23-comp), inner sep=2mm](T2) {};


  \node[lab,below = .3 of T2] (time-2) {$t_{(2)}$};

  \foreach \angle in {105, 155, ..., 255}{
  \draw[arrow] (selected1.east) -- (T2.\angle);
  \draw[arrow] (selected2.east) -- (T2.\angle);
  \draw[arrow] (selected3.east) -- (T2.\angle);
  }


  \node[atom, fill=blue] at (clust21-atom4) (delegate21) {};
  \node[atom, fill=blue] at (clust22-atom1) (delegate22) {};
  \node[atom, fill=blue] at (clust23-atom11) (delegate23) {};

  \node[atom, fill=blue, right = 3 of delegate21]  (selected21) {};
  \node[atom, fill=blue] at (selected21|-delegate22)  (selected22) {};
  \node[atom, fill=blue] at (selected21|-delegate23)  (selected23) {};

  \draw[arrow, dashed] (delegate21) -- node[pos=0.65, above, font=\bfseries]  {\scriptsize delegate} 
          node[pos=0.65, below, font=\bfseries]  {\scriptsize of $\mathcal{S}_{(2)}^1$} 
          (selected21);
  \draw[arrow, dashed] (delegate22) -- node[pos=0.65, above, font=\bfseries]  {\scriptsize delegate} 
          node[pos=0.65, below, font=\bfseries]  {\scriptsize of $\mathcal{S}_{(2)}^2$} 
          (selected22);
  \draw[arrow, dashed] (delegate23) -- node[pos=0.65, above, font=\bfseries]  {\scriptsize delegate} 
          node[pos=0.65, below, font=\bfseries]  {\scriptsize of $\mathcal{S}_{(2)}^3$} 
          (selected23);

  \node[lab, above = .3 of T1] (text-clust) {Partitioning};
  \node[lab] at (start|-text-clust) (text-start) {First\\replication};
  \path (text-start) -- (text-clust) node[midway ,lab] () {Cloning\\and\\Simulation};  
  \node[lab] at (selected1|-text-clust) (text-select) {Selection};
  \node[lab, above = .3 of T2] (text-clust2) {Partitioning};
  \path (text-select) -- (text-clust2) node[midway, lab] () {Cloning\\and\\Simulation};


\end{tikzpicture} }
    \caption{Illustration of the first and second iterations of SPSC, starting from a single initial state. }
    \label{fig:weight_proc}
    \end{figure}
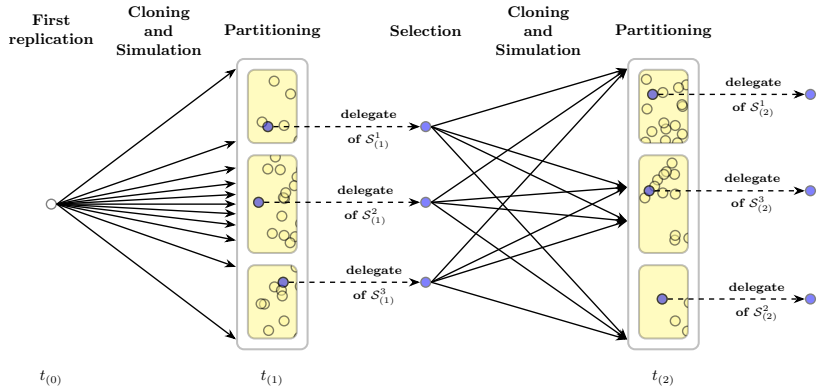

	Inspired by the  decomposition \eqref{eq:full_decomp}, we split the time interval $[0, T]$ into $m$ pieces:  $[t_{(0)},t_{(1)}], [t_{(1)},t_{(2)}], ..., [t_{(m-1)}, t_{(m)}]$ where $t_{(0)}=0 < t_{(1)} < ... < t_{(m-1)} < t_{(m)} = T$. Then, for each interval $[t_{(i)}, t_{(i+1)}]$ the following steps are applied (\emph{c.f.} Figures \ref{fig:diag_proc} and  \ref{fig:weight_proc}):
	
    \begin{description}
    \item[{Simulation }] Simulate $N$ replications from $t_{(i)}$ to $t_{(i+1)}$, $i\in\{0,\ldots,m-1\}$, where $N$ corresponds to the number of replications we authorize for the simulation.
    \item[{Partitioning }] At time $t_{(i+1)}$, form a partition of the space of observables of these $N$ replications.
    This can be done by applying a clustering algorithm. 
    \item[{Selection }] Choose one or multiple representative replications (which we call delegates) from each partition and discard the other replications.
    \item[{Cloning}] Clone the selected delegates to obtain $N$ replications in total.
    
    \end{description}
	
    Once the iterations are finished, we have created for each $t_{(i)}$ a partition $\mathscr{P}_{(i)}$ for the state space of $\boldsymbol{X}_{t_{(i)}}$. For an element $\mathcal{S}_{(i)} \in \mathscr{P}_{(i)}$ of the partition at time step $t_{(i)}$, after the selection and cloning steps, it has $n_i$ cloned replications. Besides, among these $n_i$ cloned replications, after evolving to the next time step $t_{(i+1)}$, some of them ($n_{i+1}$ replications)  belong to some element $\mathcal{S}_{(i+1)} \in \mathscr{P}_{(i+1)}$. We propose to use the numbers $n_i$ and $n_{i+1}$  to approximate the conditional probability $\mathbb{P}(\boldsymbol{X}_{t_{(i+1)}} \in \mathcal{S}_{(i+1)} | \boldsymbol{X}_{t_{(i)}} \in \mathcal{S}_{(i)})$ by:
    \begin{align}
    \label{eq:esti}
        \hat{P}(\boldsymbol{X}_{t_{(i+1)}} \in \mathcal{S}_{(i+1)} | \boldsymbol{X}_{t_{(i)}} \in \mathcal{S}_{(i)}) = \frac{n_{i+1}}{n_i}
    \end{align}
    Finally, we define an estimator $\hat{\theta}_{\mathcal{S}}$ for $\mathbb{P}(\boldsymbol{X}_{T} \in \mathcal{S})$ using a similar decomposition as that of Equation (\ref{eq:full_decomp}), based on time steps $t_{(i)}$ and \eqref{eq:esti}:
    
    \begin{align}
	    \hat{\theta}_{\mathcal{S}} = \sum_{
	        \substack{\mathcal{S}_{(m-1)} \in \mathscr{P}_{(m-1)}\\ 
	                  \vdots                                     \\
	                  \mathcal{S}_{(1)} \in \mathscr{P}_{(1)}
	        }} \prod_{i=0}^{m-1}
	    \hat{P}(\boldsymbol{X}_{t_{(i+1)}} \in \mathcal{S}_{(i+1)} | \boldsymbol{X}_{t_{(i)}} \in \mathcal{S}_{(i)})
	\end{align}
		where  
	$\mathcal{S}_{(m)} =\mathcal{S}$ and $\mathcal{S}_{(0)}=\mathcal{S}_{0}$. 

    \subsection{Implementation}
    
    We describe here a simple implementation of SPSC used in the experiment in Section \ref{sec:experiences}: 
    \begin{description}
    \item[Simulation] No special action is taken in this step.
    \item[Partitioning] $N$ replications provide $N$ instances of observables. To form a partition in the space of observables, we can take advantage of existing unsupervised learning algorithm which can separate instances by multiple subgroups. The well-known clustering process \emph{kmeans} has been chosen to fulfill the task. The number of cluster $k$ is preset to 15.
    \item[Selection] From any element $\mathcal{S}_{(i)} \in \mathscr{P}_{(i)}$ of an intermediate partition, we select the replication which is the nearest to the center using euclidean distance on the space of observables.
    \item[Cloning] After partitioning and selection, $k$ delegates are obtained to be cloned. To come back to $N$ replications in total, we clone each delegate $\lfloor \frac{N}{k}\rfloor$ times. If $k$ does not divide $N$, we select randomly the remainder number $N - k*\lfloor \frac{N}{k}\rfloor$ of delegates to produce one more clone per selected delegate.
    \end{description}    
    The time interval $[0, T]$ is homogeneously split into $m=5$ pieces (\emph{i.e.} 
    $\forall i \in \{0,1, \dots,5\}$, 
    $t_{(i)}=i\times\frac{T}{5}$).

\section{Experiment} 
    \label{sec:experiences}
	\label{sec:PPM}
	Let us take the prey predator model mentioned previously in Section \ref{sec:MC} as an  example. As this model is well-known and well-studied, we can give some possible solutions before launching simulations:
	\begin{description}
        \item[\extall:] Extinction of preys and predators, only vegetation remains.
        \item[\extpred:] Predators go extinct, preys live without nature enemy's harass.
        \item[\happyend:] All species survive and form a stable ecosystem. 
    \end{description}
	Now the question is, for an arbitrary parameter set and initial condition, what is the probability of these solutions at a given time step (\emph{e.g.} $T=1000$)? To answer this question, the MC approach recalled in Section \ref{sec:MC} is generally used.  
	
    
	In the following, we compare the performances of MC and SPSC. 
	The validation is done by comparing the outputs of these methods with the same limited number of replications $N=50$. By repeating the simulations 1000 times, we will be able to compare statistically the results obtained by MC and SPSC. Two performance measures are considered here
	: 1)~ The detection rate of a specific solution $\mathcal{S}$. 2)~The precision of the probability estimator for a specific solution. Before evaluating these performance measures, we have done 30000 replications using MC in order to provide reference values for the comparisons: 
	\begin{equation}
    \label{eq:proba_ref}
    P_{ref}(\extall) \approx 0.0065, P_{ref}(\extpred) \approx 0.0126, P_{ref}(\happyend) \approx 0.981
    \end{equation}

	
	The detection rates obtained with MC and SPSC policies, \emph{i.e}, the capacity of identifying a specific solution, from N = 50 replications are summed up in Table~\ref{tab:detec_comp}. 
	The first three columns indicate the detection rate of single solutions and the last column indicate the detection rate for the three solutions simultaneously. We can then deduce that SPSC explores more efficiently the solution space.
	\begin{table}[htbp]
	\caption{Detection rates obtained with MC and SPSC when launching 50 replications.}
    \label{tab:detec_comp}
    \centering
    \begin{tabular}{@{}r@{\hspace{.4cm}}c@{\hspace{.4cm}}c@{\hspace{.4cm}}c@{\hspace{.4cm}}c@{}}
    \toprule
              & \extall{} & \extpred{} & \happyend{} & \extall{}, \extpred{} and \happyend{} \\ \midrule
    MC        & 0.236 & 0.455 & 1     & 0.102                 \\
    SPSC    & 0.332 & 0.617 & 1     & 0.203                 \\
     \bottomrule
    \end{tabular}
    \end{table}
	
\begin{figure}[htbp]
    \centering
    \includegraphics[width=.9\linewidth]{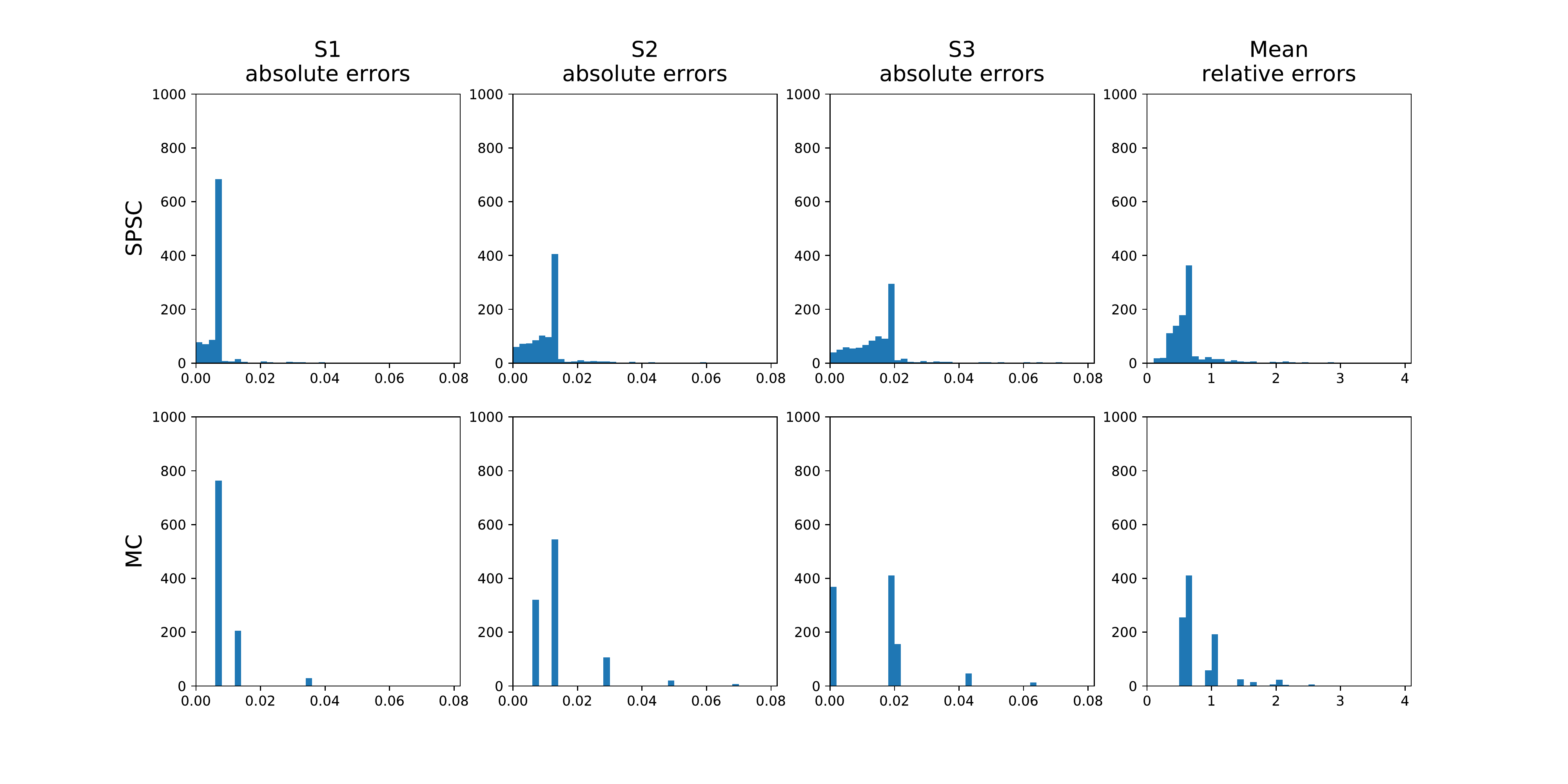}
    \caption{Comparison of errors}
    \label{fig:comp_results}
    \end{figure}

	To evaluate the precision of the probability estimator,  the absolute error between the probability estimator outcomes and the references is computed:
	\begin{align}
    Err(\mathcal{S}_i) = | \hat{P}(\mathcal{S}_i) - P_{ref}(\mathcal{S}_i) |.
    \end{align}
    Furthermore, to gain an entire vision on the three solutions simultaneously, we consider also the mean of three solutions relative errors:
    \begin{align}
    \overline{Err} =  \sum_{1 \leq i \leq 3} |\frac{\hat{P}(\mathcal{S}_i) - P_{ref}(\mathcal{S}_i)}{P_{ref}(\mathcal{S}_i)} |.
    \end{align} 
    Histograms of these errors for the policies SPSC and MC are shown in Fig.~\ref{fig:comp_results}.
	
	We can  notice that the distribution of errors are not normal nor symmetric. Thus, to compare the errors from MC and SPSC policies, the Wilcoxon-Mann-Whitney test is applied with the threshold $\alpha=0.05$. The test results are presented in Table~\ref{tab:WilcoxPP} with p-value and alternative hypotheses, we can then conclude that SPSC yields better probability estimates for each solution than MC. 
	\begin{table}[htbp]
    \caption{Hypothesis and p-value given by Wilcoxon-Mann-Whitney test. }
    \label{tab:WilcoxPP}
    \centering
    \begin{tabular}{cccc}
    \toprule
    Target solution \ \ & Alternative hypotheses \ \ & p-value & Conclusion\\
    \midrule
     \extall & $Err_{SPSC} < Err_{MC}$ & \small\texttt{2.2e-16} & $Err_{SPSC}< Err_{MC}$ \\
     \extpred & $Err_{SPSC} < Err_{MC}$ & \small\texttt{3.163e-16} & $Err_{SPSC} < Err_{MC}$\\
    \happyend & $Err_{SPSC} < Err_{MC}$ & \small\texttt{9.849e-09} & $Err_{SPSC} < Err_{MC}$\\
    \extall{}, \extpred{} and \happyend{} & $\overline{Err}_{SPSC} < \overline{Err}_{MC}$ & \small\texttt{2.2e-16} &  $\overline{Err}_{SPSC} < \overline{Err}_{MC}$\\
    \bottomrule
    \end{tabular}
    \end{table}

\section{Conclusions and perspectives}
\label{sec:conclusion}
We have introduced a generic policy called SPSC for executing stochastic simulations that deals with the weakness of MC when the number of replications is limited. It treats simulations as black boxes and therefore, does not rely upon \emph{a priori} knowledge. We have also presented a simple implementation of SPSC and run it on a classic stochastic MABS model. By comparing the results obtained with SPSC and with MC, we can conclude that SPSC gives a better solution probability estimation and can reveal more different solutions than MC.  



The first perspectives of this work are related to the impact of the parameters ($N$, $k$, etc.), the partitioning algorithm as well as the selection and cloning strategies on the performance. For instance, instead of having the same number of clones for each delegate, we could clone more the delegates from small partitions and less the delegates from large partitions. 
Theoretical properties of the proposed solution as well as its interest for multimodal transport simulation  will also be investigated. 


Moreover, since we deal with small sample size at each intermediate time step, we could take advantage of modern tools \cite{kanjanatarakul2016prediction} for statistical inference to compute the estimator $\hat{\theta}_{\mathcal{S}}$.

\section*{Acknowledgement} The ELSAT2020 project is co-financed by the European Union with the European Regional Development Fund, the French state and the Hauts de France Region Council.

    \bibliographystyle{splncs04}
    \bibliography{Biblio}
\end{document}